\documentclass[twocolumn,aps,prl,showpacs,amsmath,amssymb]{revtex4}

\usepackage{graphicx}
\usepackage{dcolumn}
\usepackage{bm}

\begin{document}

\title{Electronic Phase Diagram of High-$T_c$ Cuprate Superconductors 
from a Mapping of the In-Plane Resistivity Curvature}

\author{Yoichi Ando}
 \email{ando@criepi.denken.or.jp}
\author{Seiki Komiya}
\author{Kouji Segawa}
\author{S. Ono}
\author{Y. Kurita}
\affiliation{Central Research Institute of Electric Power Industry, 
Komae, Tokyo 201-8511, Japan.}
\date{\today}

\begin{abstract}

We propose that Resistivity Curvature Mapping (RCM) based on the
in-plane resistivity data is a useful way to objectively draw an
electronic phase diagrams of high-$T_c$ cuprates, where various {\it
crossovers} are important. In particular, the pseudogap crossover line
can be conveniently determined by RCM. We show experimental phase
diagrams obtained by RCM for Bi$_2$Sr$_{2-z}$La$_z$CuO$_{6+\delta}$,
La$_{2-x}$Sr$_x$CuO$_4$, and YBa$_2$Cu$_3$O$_y$, and demonstrate the
universal nature of the pseudogap crossover. Intriguingly, the
electronic crossover near optimum doping depicted by RCM appears to
occur rather abruptly, suggesting that the quantum critical regime, if
exists, must be very narrow. 

\end{abstract}


\maketitle

Elucidating the electronic phase diagram of high-$T_c$ cuprates in the
temperature ($T$) vs doping ($p$) plane is an important experimental
step towards understanding the high-$T_c$ superconductivity. When there
is a clear phase transition, such as the superconducting transition or
the N\'eel transition, it is easy to locate the transition line in the
$T$ vs $p$ plane. However, when there is a {\it crossover} in the
electronic properties, it becomes non-trivial to define the crossover
and to locate the crossover line. In the case of cuprates, there is an
important crossover line in the phase diagram, that is, the crossover to
the pseudogap phase within the normal state \cite{Timusk,Orenstein};
also, there is another putative crossover from a non-Fermi liquid to a
Fermi liquid in the overdoped regime \cite{Orenstein}. Because of the
lack of a universal and objective way to determine these crossover
lines, the phase diagram of the cuprates has been rather loosely
discussed in the past \cite{Orenstein,Anderson,Batlogg,Kivelson} and it
is not known to what extent the putative phase diagram is generic.

In this Letter, we propose a novel way to objectively draw the $T$ vs
$p$ phase diagram based on the in-plane resistivity ($\rho_{ab}$) data.
We call this new method ``Resistivity Curvature Mapping" (RCM), because
the curvature (second derivative) of the $\rho_{ab}(T)$ data is mapped
onto the $T$ vs $p$ plane, graphically showing how the behavior of
$\rho_{ab}$ changes as a function of temperature and doping reflecting
the changes in the underlying electronic phase. The actual phase
diagrams obtained by RCM are presented for three hole-doped cuprates,
Bi$_2$Sr$_{2-z}$La$_z$CuO$_{6+\delta}$ (BSLCO), La$_{2-x}$Sr$_x$CuO$_4$
(LSCO), and YBa$_2$Cu$_3$O$_y$ (YBCO), in all of which the hole doping
can be widely changed from underdoped to overdoped regimes. The phase
diagrams of the three systems demonstrate that the pseudogap crossover
line, which can be conveniently determined by RCM, shows a linear doping
dependence (in the superconducting doping regime) and is terminated near
optimum doping. Furthermore, our data demonstrate that the well-known
``$T$-linear" resistivity is observed in a very limited region of the
phase diagram near optimum doping, which suggests that the
quantum-critical non-Fermi-liquid regime \cite{Orenstein}, if exists,
must be much narrower than is often conjectured
\cite{Orenstein,Sachdev}.

The single crystals of BSLCO, LSCO, and YBCO are grown by a
floating-zone technique \cite{Ono}, a traveling-solvent floating-zone
technique \cite{Komiya}, and a flux method \cite{Segawa}, respectively.
For BSLCO, we first determined the actual La content, $z$, of
representative samples by the inductively-coupled plasma atomic-emission
spectroscopy (ICP-AES) and established a relationship between $z$ and
the $c$-axis lattice constant $c_0$; we then measure $c_0$ of all the
samples to determine actual $z$. Since the relation between $z$ and $p$
(hole doping per Cu) has been sorted out in our previous study
\cite{Hanaki}, here we show just the $p$ values for our BSLCO samples
for simplicity. In the case of LSCO, one can safely assume that the $p$
value is equal to $x$ (Sr content), as long as the samples are carefully
annealed to remove excess oxygen or oxygen vacancies, which is what we
always do for our LSCO crystals \cite{Komiya}. On the other hand, it is
difficult to determine the exact hole doping $p$ in the CuO$_2$ planes
of YBCO \cite{Segawa_Hall}, so here we just show the oxygen content $y$,
which is measured by iodometry and is the control parameter of the
doping. We have established a technique to reliably and reproducibly
anneal the YBCO crystals to tune $y$ to a target value \cite{Segawa}; we
always quench the sample to room temperature at the end of the annealing
to avoid the long-range oxygen ordering phenomena, which often
complicate the physics of YBCO \cite{Ando_MOS}. Our previous study
indicated that the hole doping $p$ changes smoothly with $y$ (without
showing a plateau near $y \sim$ 6.6) in our quenched samples
\cite{Segawa,Segawa_Hall}. All the YBCO samples are untwinned crystals
and the $a$-axis resistivity $\rho_a$ is measured to avoid direct
contributions from the CuO chains, which run along the $b$-axis. The
resistivity is measured with a standard ac four-probe technique
\cite{mobility}.

\begin{figure} 
\includegraphics[clip,width=7.0cm]{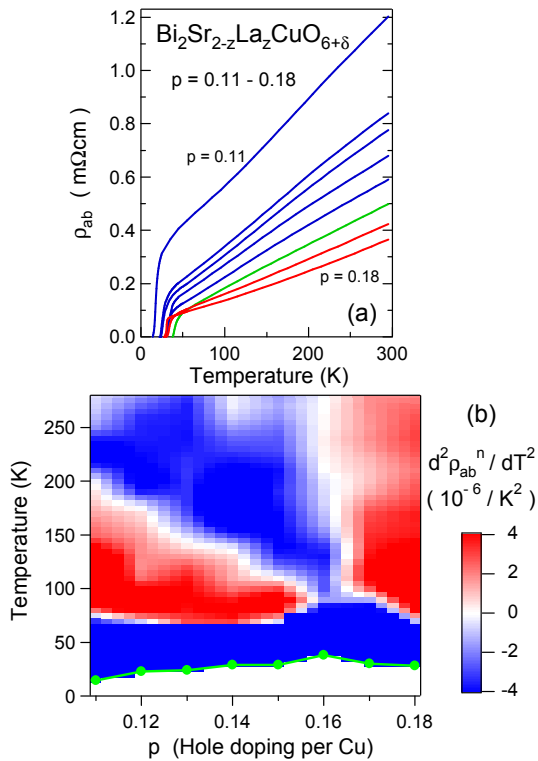}
\caption{(a) $\rho_{ab}(T)$ data of BSLCO for $p$ = 0.11--0.18
at 0.01 intervals; the data for underdoped (overdoped) samples are shown 
in blue (red), while that for optimum doping is shown in green. 
(b) Resistivity curvature mapping, namely, a false color mapping of 
$d^2\rho_{ab}^{\rm n}/dT^2$ in the $T$ vs $p$ plane, for BSLCO.
The solid green circles show $T_c$'s for the measured compositions.} 
\end{figure}

Figure 1(a) shows the $\rho_{ab}(T)$ data of BSLCO for $p$ = 0.11 --
0.18 at 0.01 intervals; note that optimum doping of this system
corresponds to $p$ = 0.16 where the zero-resistivity $T_c$ is 36--38 K
\cite{Ono_MI,Hanaki}. To compare the $\rho_{ab}(T)$ behavior for
different dopings, we normalize $\rho_{ab}$ by its 300-K value to obtain
$\rho_{ab}^{\rm n}$ [$= \rho_{ab}/\rho_{ab}(300{\rm K})$]. By
calculating the second derivative of $\rho_{ab}^{\rm n}(T)$,
$d^2\rho_{ab}^{\rm n}/dT^2$, for each doping (wherein we employ modest
numerical smoothing \cite{note}) and linearly interpolating the results,
we can draw a false color mapping of $d^2\rho_{ab}^{\rm n}/dT^2$ in the
$T$ vs $p$ plane as shown in Fig. 1(b), where the red (blue) color
corresponds to positive (negative) curvature, while the white color
means that the curvature is essentially zero. Note that when
$\rho_{ab}(T)$ changes linearly with $T$, the curvature of
$\rho_{ab}(T)$ is zero; thus, the vertical white band apparent in Fig.
1(b) at $p \simeq$ 0.16 means that the $T$-linear resistivity is
observed there. Furthermore, the vertical red region in Fig. 1(b) for $p
\agt$ 0.17 graphically demonstrates that as soon as BSLCO is overdoped
the $\rho_{ab}(T)$ behavior loses the $T$-linearity; actually, we have
shown \cite{Murayama} that $\rho_{ab}(T)$ of the overdoped BSLCO samples
are well fitted with $\sim T^{\alpha}$ with $\alpha > 1$. 

More importantly, Fig. 1(b) also depicts the effect of the pseudogap
opening in underdoped samples: It is known \cite{Murayama,Ito} that the
pseudogap tends to cause $\rho_{ab}(T)$ to show an ``S-shape" behavior,
which has been interpreted to be due to a rapid reduction in the
inelastic scattering rate of the electrons in the pseudogap phase upon
lowering temperature \cite{Ito}. Although it was recently proposed
\cite{Hall} that this S-shape may better be viewed in terms of a gradual
participation of the quasiparticles near $(\pi,0)$ of the Brillouin zone
with increasing temperature, in any case the partial destruction of the
Fermi surface \cite{Norman} is responsible for the peculiar
$\rho_{ab}(T)$ behavior in the pseudogap phase. If one accepts the
S-shape to be a signature of the pseudogap in the dc transport, then
there is a unique and unambiguous temperature which characterize the
S-shape, that is, the inflection point. It is thus natural to consider
the inflection point in the $\rho_{ab}(T)$ curve as a ``characteristic
temperature" of the pseudogap, $T_{pg}$, though it may not denote the
``onset" of the pseudogap. In our RCM the inflection point shows up in
white, so the diagonal white band that vertically separates the blue and
red regions in the underdoped regime ($p < 0.16$) signifies how the
characteristic temperature for the pseudogap crossover changes with
doping. It is intriguing to see that the pseudogap boundary signified by
$T_{pg}$ changes essentially linearly with doping and is terminated near
optimum doping.

In the literature, the pseudogap temperature $T^*$ has often been
extracted from the $\rho_{ab}(T)$ data by identifying the temperature
below which $\rho_{ab}$ deviates downwardly from the high-temperature
$T$-linear behavior \cite{Ito,Watanabe}. However, there are two major
shortcomings in this approach. First, in many cases there is not a
wide-enough temperature range where the ``high-temperature $T$-linear
behavior" is unambiguously determined; in fact, our data in Fig. 1(b)
demonstrate that there is no well-defined $T$-linear region at high
temperature (which should identify itself as a horizontal white band at
the top of the diagram) in underdoped BSLCO. Second, it is difficult to
objectively identify a ``deviation point" when the deviation occurs
gradually and continuously. These shortcomings have been part of the
reasons why $T^*$ has been only loosely discussed and its generic
behavior is not agreed upon. Furthermore, since the paraconductivity due
to the superconducting fluctuations (SCF) also reduces $\rho_{ab}$ and
leads to a negative curvature in $\rho_{ab}(T)$, it is sometimes
difficult to judge whether the deviation from the high-temperature
$T$-linear behavior is due to the pseudogap or the SCF, particularly
when the SCF can start from quite a high temperature \cite{YBCO_MR}. [In
Fig. 1(b), the blue region immediately above $T_c$ is where the SCF is
dominant.] An obvious merit of the present definition of $T_{pg}$ is
that those ambiguities can naturally be avoided.

\begin{figure} 
\includegraphics[clip,width=7.5cm]{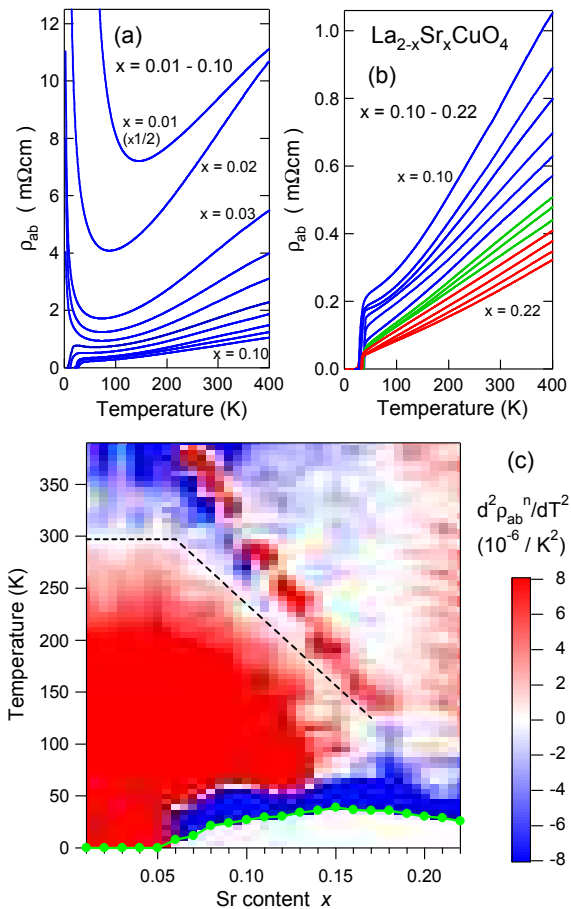}
\caption{(a,b) $\rho_{ab}(T)$ data of LSCO for $x$ = 0.01--0.22
at 0.01 intervals. 
(c) Electronic phase diagram depicted by RCM for LSCO; 
here $\rho_{ab}^{\rm n} = \rho_{ab}/\rho_{ab}(400{\rm K})$.
The dashed line is a guide to the eyes to emphasize $T_{pg}$ and the 
solid green circles show $T_c$'s for the measured compositions.} 
\end{figure}

Next we discuss LSCO, for which the raw $\rho_{ab}(T)$ data are shown in
Figs. 2(a-b). For clarity, we use the nominal $x$ values for the LSCO
samples here, but we confirmed that the actual Sr contents measured by
ICP-AES agree with the nominal $x$ values within an error of less than
5\%. The RCM plot for LSCO [Fig. 2(c)] is a bit complicated due to the
existence of the structural phase transition from the high-temperature
tetragonal phase to the low-temperature orthorhombic phase, which causes
a weak kink in the $\rho_{ab}(T)$ data; the diagonal red band that ends
at $x$ = 0.18 signifies this transition, whose position is consistent
with the data in the literature \cite{Kastner}. Apart from this
structural transition, one can see that the phase diagram of LSCO
depicted by RCM is very similar to that of BSLCO in several
respects: First, the $T$-linear resistivity (vertical white band) is
observed only near optimum doping ($x \simeq$ 0.16 -- 0.18). Second,
the vertical red region for $p \agt$ 0.19 demonstrates that the
$\rho_{ab}(T)$ behavior becomes positively curved in the overdoped
regime. Third, $T_{pg}$ (marked by a dashed line) changes approximately
linearly with $x$ for $x \ge 0.06$ and is terminated near optimum
doping; incidentally, it is intriguing to see that $T_{pg}$ 
saturates in the non-superconducting regime ($x < 0.06$) and that the
saturated value of $T_{pg}$ is close to the N\'eel temperature for $x$ =
0 ($\sim$300 K) \cite{Kastner}.

\begin{figure} 
\includegraphics[clip,width=8cm]{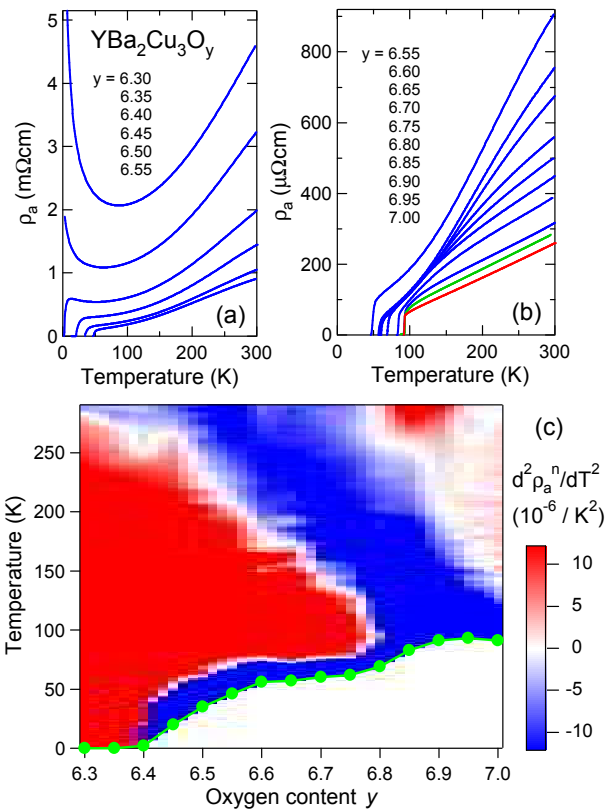}
\caption{(a,b) $\rho_{a}(T)$ data of YBCO for $y$ = 6.30--7.00
at 0.05 intervals. 
(c) Electronic phase diagram depicted by RCM for YBCO, where the solid 
green circles show $T_c$'s for the measured compositions.} 
\end{figure}

Figures 3(a-b) show the $\rho_a(T)$ data for YBCO, and the RCM plot is
shown in Fig. 3(c). One can easily see that the phase diagram of YBCO
depicted in Fig. 3(c) is quite similar to those of BSLCO and LSCO in
that (1) the $T$-linear resistivity is observed only near optimum doping
({\it i.e.}, $y \simeq$ 6.95), and (2) $T_{pg}$ changes approximately
linearly with $y$ in the superconducting regime and tends to saturate in
the antiferromagnetic regime. In addition, one can see that
$\rho_{a}(T)$ becomes slightly positively curved in the overdoped
regime, which is recognized by the faint red color at $y$ = 7.00 for $T
>$ 150 K. However, Fig. 3(c) also shows a departure from the universal
phase diagram suggested by BSLCO and LSCO in two aspects: (1) $T_{pg}$
is terminated at $y \simeq$ 6.8, which is near optimum doping but is in
the underdoped regime, and (2) the high-temperature behavior at $6.80
\alt y \alt 6.90$ is complicated. (The red blob at the top of the
diagram for $6.85 \le y \le 6.90$ is due to a slight curving of the
$\rho_{a}(T)$ data near 300 K at these dopings, which we confirm to be
very reproducible; this is due to the oxygen motion in the Cu-O chains 
\cite{oxygen}.) Phenomenologically, it appears that these peculiarities are
related to the fact that the $T_c$-vs-$y$ diagram of YBCO [see the green
symbols in Fig. 3(c)] shows two plateaus at $\sim$60 K and $\sim$90 K,
the former called the 60-K phase and its origin is still under debate
\cite{Segawa}; clearly, the phase diagram is more ordinary in the 60-K
phase and below ($y \le$ 6.80), but becomes peculiar near the 90-K
phase. One interesting possibility is that the phase diagram for $y \le$
6.80 reflects the physics of the CuO$_2$ planes alone, while in the 90-K
phase the Cu-O chains introduce additional complications. In any case,
the slight nonuniversality in YBCO is most likely related to its
structural peculiarity.

The above results demonstrate that the cuprate phase diagram is
surprisingly universal unless some additional feature (such as the Cu-O
chains in YBCO) adds another layer of complications. In particular, the
fact that the $T$-linear resistivity is observed only near optimum
doping and $\rho_{ab}(T)$ becomes positively curved as soon as the
system is overdoped is commonly observed in the three cuprate systems
studied here. In this regard, our data do not give strong support to the
notion that the $T$-linear resistivity is associated with a quantum
critical region (where the physics is scaled by a single energy scale
$T$ \cite{Sachdev}); based on our data, one should probably conclude
that the quantum critical region, if exists, must be much narrower in
cuprates than is observed in other systems \cite{Custers,Perry}, where
it clearly fans out with increasing $T$ like an inverse triangle. It
seems that our data (particularly those of BSLCO) would rather suggest
that there are simply two different electronic states for underdoped and
overdoped regimes, and the $T$-linear resistivity appears to be a
property of a singular electronic state realized only at optimum doping.
It is useful to note that such an abrupt nature of the crossover at
optimum doping might be related to a change in the underlying
Fermi-surface state, which was recently shown to occur at optimum doping
in BSLCO \cite{Balakirev}.

Finally, it is useful to note that in all three cuprates the pseudogap
crossover line $T_{pg}(p)$ is terminated somewhere near optimum doping
and it does not extend into the overdoped regime. This might appear to
be inconsistent with the results of the angle-resolved photoemission
spectroscopy (ARPES) \cite{Shen} or the scanning tunneling microscope
(STM) \cite{Kugler} that observed a pseudogap in overdoped samples, but
such apparent inconsistency is probably a manifestation of the fact that
the pseudogap in the cuprates has two different origins
\cite{Lavrov,Markiewicz} and only one of them is responsible for the
S-shape in $\rho_{ab}(T)$. In this regard, it is useful to note that the
SCF are probably related to various pseudogap features that occur rather
close to $T_c$ \cite{Lavrov,Markiewicz,Wang}, and in this sense the blue
region immediately above $T_c$ in our RCM plots can be considered to be
representative of the pseudogap of a different origin; this
interpretation is actually consistent with the phase diagram of BSLCO
depicted by the $c$-axis magnetoresistance \cite{Lavrov}. In passing, we
note that the pseudogap is known to be most directly probed in the
$c$-axis properties \cite{Timusk,Orenstein,Lavrov} and $T_{pg}$
determined here from RCM is rather indirectly reflecting the pseudogap,
probably giving a lower measure of its development.

In summary, we demonstrate that the Resistivity Curvature Mapping
(RCM) offers a powerful tool to draw an electronic phase diagram of
high-$T_c$ cuprates based on the resistivity data. The biggest merit of
this method is that it allows one to objectively draw the pseudogap
crossover line $T_{pg}(p)$. It also allows one to see where in the phase
diagram the $T$-linear resistivity is observed. The RCM-derived phase
diagrams we present for BSLCO, LSCO, and YBCO demonstrate that the
essential feature of the phase diagram is surprisingly universal and
that the electronic crossover near optimum doping occurs rather abruptly.

We thank A. N. Lavrov for helpful discussions.


\begin{thebibliography}{}
\bibitem{note}
At each $T$, we fit the raw data in a window $T \pm \Delta T$ with a
polynomial of order 2 -- 4 and take its second derivative ($\Delta T$ is
typically 5 K); the resulting RCM is essentially the same for $\Delta T$
of 2 -- 10 K, except that a sharp feature gets gradually smeared with
increasing $\Delta T$.
\end{thebibliography}
\end{document}